\documentstyle[11pt,newpasp_sb,twoside,epsf]{article}
\def\edcomment#1{\iffalse\marginpar{\raggedright\sl#1\/}\else\relax\fi}
\reversemarginpar

\begin{document}
\title{Submillimeter Array observations of ISM in starburst galaxies}
 \author{Kazushi Sakamoto$^1$, 
 Satoki Matsushita$^2$, 
 Alison B. Peck$^1$, 
 Rui-Qing Mao$^{2,3}$, 
 Martina C. Wiedner$^{1,4}$,
 Daisuke Iono$^1$, 
 Cheng-Yu Kuo$^2$,
 and SMA team$^{1,2}$}
\affil{1. Submillimeter Array, Harvard-Smithsonian Center for Astrophysics, Hilo, HI, 96721-0824, U.S.A.}
\affil{2. Institute of Astronomy and Astrophysics, Academia Sinica,  P.O. Box 23-104,
Taipei 106, Taiwan}
\affil{3. Purple Mountain Observatory, Nanjing 210008, China}
\affil{4. Physikalishes Institut, Univ. of K\"{o}ln, 50923, K\"{o}ln, Germany}

\begin{abstract}
The Submillimeter Array (SMA) has been under construction at the 4100 m summit of
Mauna Kea, Hawaii. 
The array is going to allow imaging of  lines
and continuum at sub-arcsecond resolution in submillimeter wavelengths.
The status of the array and the results from recent commissioning observations of
nearby galaxies are reported. 
\end{abstract}

\section{Introduction}

Molecular gas and dust in starburst galaxies are known to be warmer
and denser than those in non-starburst environments.
Emission lines from highly excited molecules and dust continuum in 
submillimeter are therefore among the best probes to study neutral ISM 
in starburst galaxies.
Detailed study of starbursts in external galaxies needs a resolution of the 
order of an arcsecond.
An interferometer is necessary to achieve this resolution in submillimeter.
However, there has not been a submillimeter interferometer since the 
pioneering JCMT--CSO interferometry (c.f. Wiedner 2002).
SMA is going to be the first full-time submillimeter interferometer with fast 
imaging capability of eight antennas. Though still in construction and
testing phase,  the array has recently started its initial observations of galaxies 
as reported here. 
Another contribution by Iono et al. in these proceedings shows preliminary results of
more distant starbursts.

\section{SMA overview}
SMA is a joint project between CfA and ASIAA to build the first
full-time interferometer with submillimeter capability at the summit of
Mauna Kea. The first SMA antenna arrived at the site in 1999, and 
the eighth antenna arrived in 2003. 
The array has been undergoing various tests for commissioning.
Fringes were first detected on the 
entire 28 baselines on Nov. 06, 2003.
The dedication of the SMA is in late November 2003.  
Figure 1 shows the SMA in fall 2003.

Table 1 summarizes SMA parameters, with some values being
those expected at the time of dedication. 
The array consists of eight 6-meter antennas each with surface accuracy of
about 12 \micron. 
The antennas can be placed on any of 24 pads that provide a maximum 
baseline of 500 m or maximum resolution of 0.5--0.1 arcsec in
200 -- 900 GHz bands.
Up to eight receivers can be installed on each antenna to cover
most atmospheric windows in submillimeter wavelengths. 
Our three initial receivers currently installed are for 
230, 345, and 690 GHz bands. 
The digital correlators provide a bandwidth of 2 GHz and
a resolution of 0.2--0.8 MHz.

The initial reduction of SMA data is made using MIR, which is an
IDL-based package kindly provided from Caltech and is under development
in collaboration with the OVRO group.
The data are then processed in reduction packages such as AIPS and MIRIAD.

\begin{figure}
\plotone{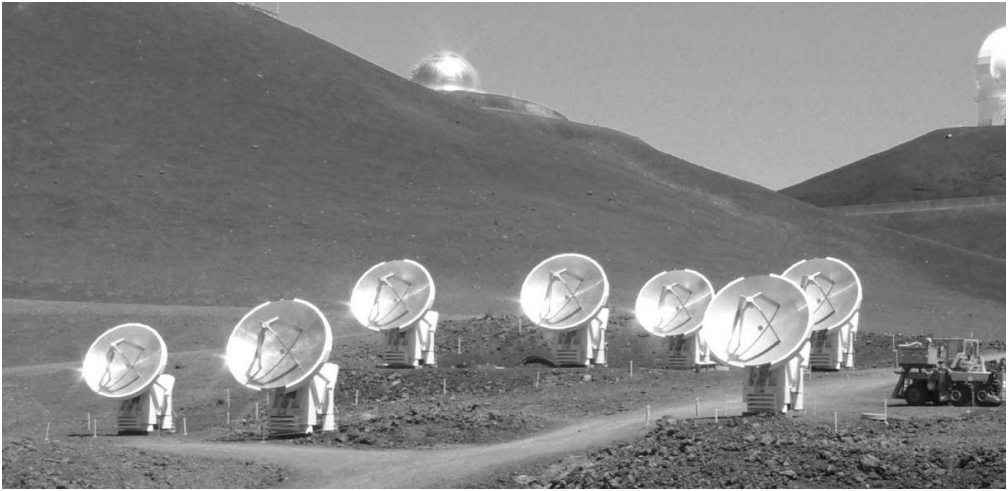}
\caption{Submillimeter Array at Mauna Kea in Sept. 2003.} 
\end{figure}

\begin{table}
\caption{Key parameters of SMA}
\begin{tabular}{lc}
\tableline
Antennas   	& Eight 6 m, rms $\approx$ 12 $\mu$m, CFRP back structure\\
Operating frequency & 200 -- 900 GHz \\
Receivers			& up to 8 per antenna, dual freq./pol. capable\\
Initial receivers 	& 230, 350, 650 GHz bands\\
Array configuration	& 24 pads on 4 Reuleaux triangles \\
Baselines		& 8 -- 508 m, 28 baselines per configuration \\
Spatial resolution & up to 0.5 -- 0.1 arcsec \\
Filed of view 	& 60 -- 14 arcsec (FWHP)\\
Bandwidth 	& 2 GHz (4 GHz  via sideband separation)\\
Spectral resolution & 0.8 -- 0.2 MHz/channel. 3k--12k channels per sideband  \\
Site			& Mauna Kea, Hawaii. 155\deg W, 20\deg N, 4100 m \\
Web site		& {\sf http://sma-www.cfa.harvard.edu }\\
\tableline
\end{tabular}
\end{table}

\section{Initial observations of nearby galaxies with SMA}

We observed the nucleus of M51 and M83 in the course of SMA commissioning.
The choice of the targets was largely determined by the timing when
the array became operational, the CO brightness of the sources, and
their narrow line widths that fit in our 300--600 MHz bandwidth at the time of the
observations.
 
 \subsection{M51}
M51 was observed in  February 2002, and was the
first galaxy observed with 5 antennas at SMA.
 Figure 2 compares the central 1 arcmin of M51 as observed in
 CO(1--0) at NMA (Sakamoto et al. 1999) and as observed in 
 CO(2--1) at SMA. 
 The two arrays provided roughly the same field of view and the same
 resolution (4\arcsec). 
The overall spiral morphology is similar, assuring the validity of the SMA
observations.
At the nucleus, however, it is immediately clear that the molecular gas in 
the central few 100 pc  is more highly excited than the gas elsewhere.
Though M51 is not a starburst galaxy, the example here demonstrates
SMA's capability to provide valuable information on the conditions of
ISM through observations of higher excitation lines.
We made further observations of the galaxy in 2003 in CO(3--2); the results
are going to be reported in Matsushita et al. (2004).

\begin{figure}
\plottwo{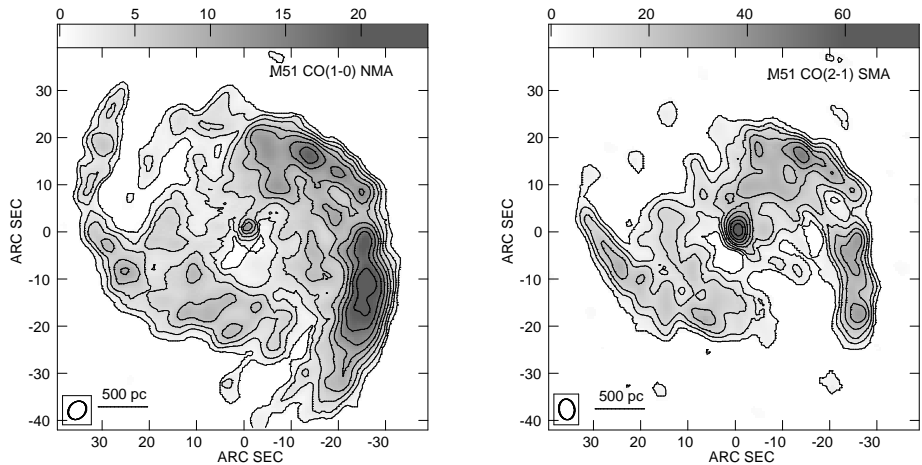}{M51_CO21SMA.eps}
\caption{Central 3 kpc of M51. Left:  CO(1--0)  (NMA; Sakamoto 
et al. 1999). 
Right: CO(2--1) from SMA commissioning observations.
Contours are at the same fraction of the peak in each map.}
\end{figure}

\subsection{M83}
The barred starburst galaxy M83 was observed in the spring of 2003 with 5 antennas.
We did the first mosaic observations at SMA in CO(2--1) in order to better observe 
the gas distribution along the stellar bar. 
We also observed CO(3--2) with a single pointing at the starburst
nucleus as one of our first CO(3--2) observations of galaxies.
Figure 3 shows the CO(2--1) mosaic and the CO(3--2) map.
The maps clearly show the typical gas morphology in barred galaxies; 
a pair of gas ridges at the leading side of the stellar bar 
and a nuclear ring of $\sim 300$ pc diameter.
The high sensitivity of SMA allowed us to detect CO(2--1) emission between the
gas ridges and the nuclear ring; the emission forms a parallelogram shape together
with the pair of gas ridges. This component is likely the `spray'
gas that traveled through a gas ridge, passed the connecting point of the ridge and 
the nuclear ring,  and is now going back to the other gas ridge.
The gas in the observed region will eventually be funneled to the nucleus to sustain the 
nuclear starburst.
The starburst in this galaxy is mainly on the `starburst arc' (Harris et al. 2001) which 
is just inside the nuclear ring of molecular gas. 
It is notable that little molecular gas is seen on the starburst arc.
M83 has two nuclei $\sim$3\arcsec\ apart (Thatte et al. 2000). 
Our data show that one of them is off the dynamical center and has high-velocity gas, which
is the peak next to the cross in the CO(3--2) map.
The dynamical perturbation from this off-center nucleus may have played a role in
triggering the starburst.
Further analysis is  reported in Sakamoto et al. (2004).

\begin{figure}
\plottwo{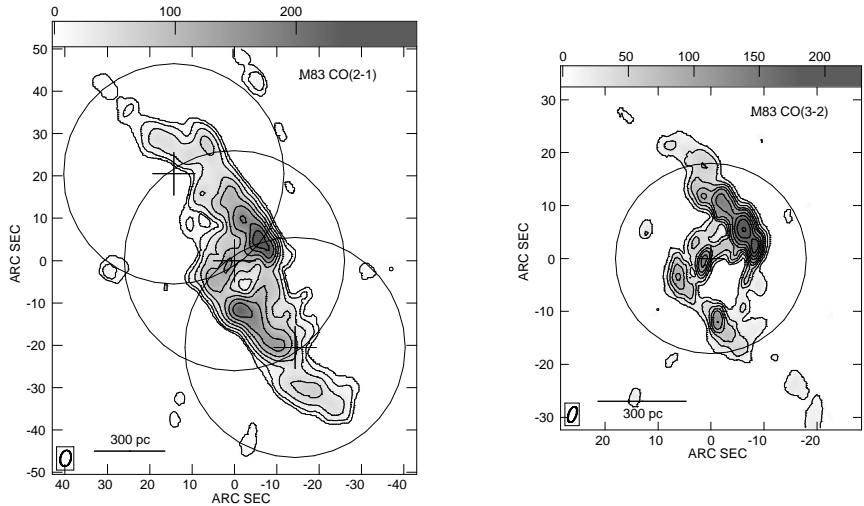}{M83_345.eps}
\caption{M83 in CO(2--1) and CO(3--2) from SMA commissioning observations. 
Fields of view and pointing positions are also shown.}
\end{figure}
	
\acknowledgements
SMA is made possible by the dedication of a large number of people. 
We are grateful to all those who contributed to the project.

\end{document}